# T2HK: J-PARC upgrade plan for future and beyond T2K

**T. Ishida, for the Hyper-Kamiokande working group**

Neutrino section, J-PARC center
203-1 Shirakata, Tokai-mura, Naka-gun, Ibaraki 319-1106, Japan

E-mail: taku.ishida@kek.jp

**Abstract**. Upgrades to the J-PARC accelerators and the neutrino experimental facility are of vital importance to the Tokai to Hyper-Kamiokande project (T2HK), which aims to explore CP asymmetry by the $\nu_e$ appearance. In this talk I present overview of the T2HK project, current status of the beam operation at J-PARC, and prospect to realize the rated 750 kW operation in coming years.

## 1. Project overview

T2K (*Tokai-to-Kamioka*) is a currently running long-baseline neutrino oscillation experiment. The baseline runs 295 km between the source of the neutrino beam at J-PARC, Tokai-mura village, and Super-Kamiokanade (Super-K), a 50,000*t*-large underground water Cherenkov detector at Kamioka mine, which serves as a far detector of neutrinos. The collaboration has recently announced the observation of $\nu_\mu$-to-$\nu_e$ appearance with significance over $7\sigma$[1]. It reveals that the search for CP violation in the lepton sector is in fact within reach of the conventional super-beam experiments. T2HK (*Tokai-to-Hyper-Kamiokande*) can be recognized as a natural extension of the technique being proven by the success of T2K. It will employ the same neutrino experimental facility at J-PARC with increased beam power, and Hyper-Kamiokande (Hyper-K), a proposed water Cherenkov detector of 1 megaton volume[2]. The primary goal of T2HK is discovery of the CP asymmetry. With its short baseline length, the CP violation effect will be more prominent than the matter effect, in comparison with the other proposed future experiments with more than 1,000km baseline length[3][4].

Figure 1 shows layout of T2HK. It has the same off-axis angle and base-line length as T2K. The candidate site of Hyper-K is located at 8 km south of Super-K, under mountains with 648 m of rock overburden. It is accessible horizontally through a mining gallery, where an ample flow (13,000m$^3$ per day) of ground water exists. The proposed detector design consists of two parallel 247.5 m-long caverns, with 48 m-wide and 54 m-high egg-shaped cross sections. Each cavity is divided into five 49.5 m-long compartments. The total (inner) volume is 0.99 (0.74) Mt, and fiducial volume is 0.56 Mt (0.056 Mt×10 compartments), 25 times larger than that of Super-K (22.5 kt). The surface of the inner detector will be covered by 99,000 20"$\phi$ PMTs with 20 % photo coverage. 25,000 of 8"$\phi$ PMT will be installed in the outer detector. We estimate 7 years for construction: 2 years to make access tunnels, 2.5 years for excavation of caverns, 2.5 years for construction. The target schedule is to start construction around Japanese fiscal year (JFY) 2016 and begin T2HK experiment around 2023.

For T2HK, upgrade of J-PARC is vitally important. LoI[2] assumes 7.5MW×year. With rated beam power (750 kW), it corresponds to 10 years (3 years with neutrino beam and 7 years with anti-neutrino beam). If systematic error is suppressed to 5 % and mass hierarchy is known, we can discover CP violation with more than 3 $\sigma$ significance in three-quarters of the domain of the CPphase $\delta$.

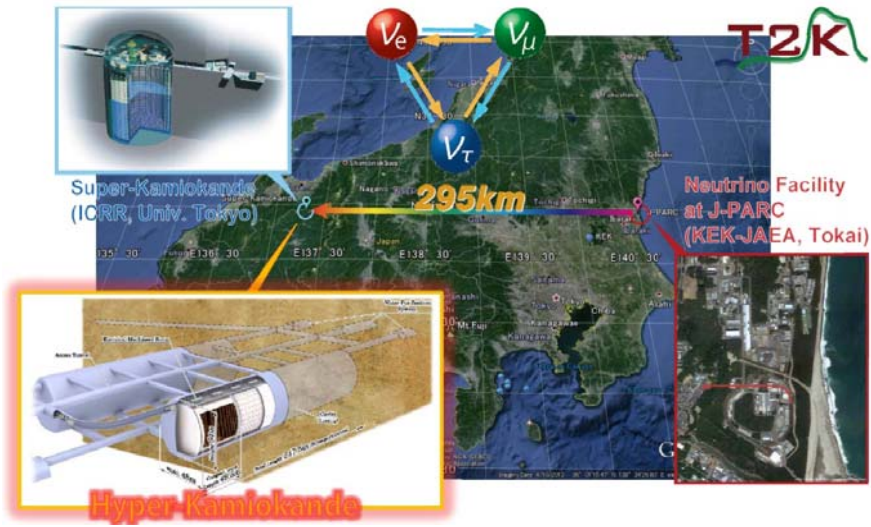

**Figure 1.** Layout of the T2HK experiment together with T2K.

## 2. J-PARC accelerator cascade and the neutrino experimental facility

J-PARC, Japan Proton Accelerator Research Complex, comprises a high intensity proton accelerator cascade and surrounding experimental facilities. The accelerator cascade consists of a normal conducting LINAC as an injection system, Rapid Cycling Synchrotron (RCS), and Main Ring Synchrotron (MR). The RCS provides a 3GeV beam to the materials and life science facility, and also plays a role as a booster synchrotron of MR. MR provides beam to the hadron hall by slow extraction, and to the neutrino experimental facility by fast extraction. Through a MR straight section, the proton beam is extracted inward to the primary beam-line, and is guided to the production target / pion-focusing horn system. The pions decay into muons and neutrinos during their flight in a 110 m-long decay volume. A beam dump consisting of graphite blocks is installed at the end of the decay volume, and muon monitors exist downstream of the beam dump to monitor the muon profile. At 280 m downstream of the target, there is a neutrino near detector complex to monitor neutrinos at production.

The target-horn system is composed of three electromagnetic horns (horn-1, -2, and -3 from upstream to downstream) and a pion production target, inserted into upstream center of the horn-1. The horns are suspended from the wall of the target station vessel, being kept under Helium atmosphere together with the decay volume, and shield blocks cover entire surface of the system for radiation protection. Each horn consists of inner and outer conductors made of aluminum alloy (A6061-T6). By applying 320 kA pulsed current (rated), a toroidal magnetic field of 2.1 Tesla at maximum is generated between the conductors to focus charged pions. During the beam operation, horns are exposed to heat load from both energy deposition by beam particles and joule heating by pulsed current. For 750 kW operation, it is estimated to be ~25 kJ total heat load per beam shot. There are water spraying nozzles assembled to the outer conductors, which spray cooling water to the inner conductor[5].

The pion production target is made of isotropic graphite rod (IG-430U) of 26 mm-$\phi \times$910 mm-L, covered by two co-axial sleeves made of the same graphite (inside) and of Titanium alloy (Ti-6Al-4V, outside), with a few mm gaps between them. During 750 kW operation, ~20 kW heat load is produced in the target, which is cooled by Helium gas flowing through the gaps. The nominal flow rate is 9 Nm$^3$/min, where the velocity of helium gas in the gaps is ~200 m/s. The maximum temperature of the graphite at thermal equilibrium is to be 740 degree C. This (rather high) temperature aims to boot up annealing effect to repair radiation damage in the graphite. We expect 0.25 displacements per atom (DPA) in one year of operation[6].

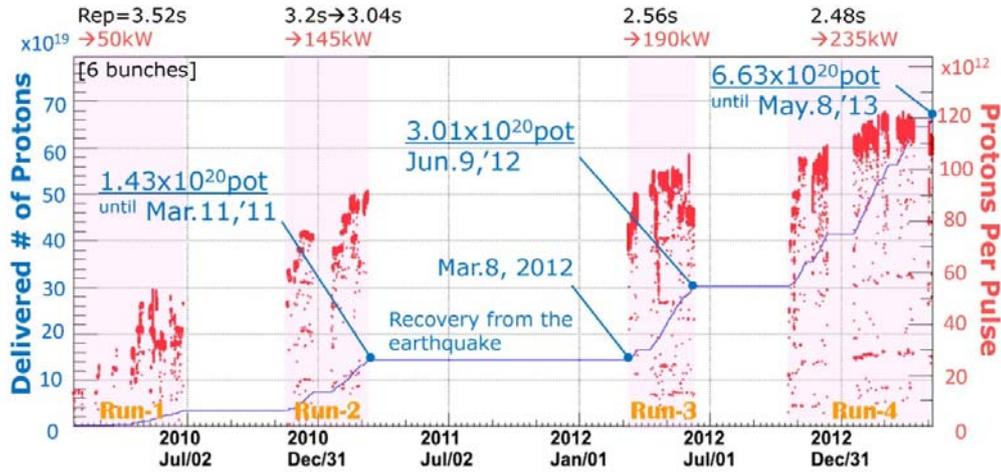

**Figure 2.** Number of protons delivered by J-PARC MR to the neutrino experimental facility.

Figure 2 shows the history of the number of protons per pulse (*ppp*) and total accumulated number of protons on target (*pot*), which are delivered from MR to the neutrino experimental facility. Beam commissioning to the neutrino experimental facility was started in April 2009, and data taking for physics run was started in early 2010 with 20 kW beam power and 3.5 sec repetition cycle. Since then, we have continuously tried to increase *ppp* and to shorten the cycle[7]. Recently, we have achieved stable operation with 220 kW beam power (at maximum 235 kW for trial) and repetition cycle of 2.48 sec. The *ppp* is more than $1.2 \times 10^{14}$ ($1.5 \times 10^{13} \times 8$ bunches), which is to be noted as the world record for synchrotrons. The accumulated *pot* is $6.63 \times 10^{20}$ and accumulated number of pulses is $1.2 \times 10^{7}$ by May, 2013. So far we succeeded to use the first target-horn system without any serious trouble. The nominal value of the horn current was set to 250 kA. Current stability was within ±5 kA (2 %) for each data taking period with respect to the average value.

## 3. Upgrade plan for J-PARC accelerators / neutrino experimental facility

For the present LINAC, beam energy is 181 MeV and peak current is 30 mA. During 2013 summer shutdown, a new accelerating structure, ACS (Annular Coupled Structure linac), is being installed at downstream of LINAC. It will recover the beam energy to 400 MeV. In 2014 summer shutdown, the front-end part, ion source and RFQ, will also be replaced to increase the peak current up to 50 mA. Even with the upgrade, *ppp* of MR will still be limited by beam loss coming from the space charge effect. For current beam energy (30 GeV), the maximum is estimated to be $2.2 \times 10^{14}$ *ppp*. With repetition rate of 0.4 Hz, the beam power will be up to 450 kW.

To achieve the rated 750 kW beam power, we aim to make beam energy higher than 30 GeV. Actually the original specification of MR adopts 50 GeV, and 750 kW beam power corresponds to $3.3 \times 10^{14}$ *ppp* with 3.5 sec cycle. However, we realized that, for beam energy more than 40 GeV, the field quality of the main magnets deteriorates due to the saturation effect. Also, the total power consumption by the magnets for 40 (50) GeV operation will be 2 (4) times larger than that for 30 GeV operation. Taking these problems into account, we decided to change our strategy: Instead of making beam energy higher, we are going to make repetition rate higher than current 0.4 Hz.

750 kW beam power is to be realized with 30 GeV beam energy, $2.0 \times 10^{14}$ *ppp* with 1.3 sec cycle. To double the rep-rate (2.5s → 1.3 s cycle), we are going to promote following mid-term plan[8]:

- Replacement of the magnet power supplies: All the main magnet power supplies will be replaced with new ones, which realize high repetition rate with energy recovery scheme using condenser banks. A new building for the power supplies is to be constructed.

- Replacement of RF cavities: New magnetic core material (FT3L), which has two times higher impedance than present ones, was developed and mass production is being started. The new RF system is designed to have more gaps, and total RF voltage will be increased from 270 kV to ~450kV.
- Upgrade of injection and extraction devices.
- Upgrade of ring collimator section: The capacity of the ring collimator section is being gradually reinforced from 450 W (~2011) to 2 kW (2012), and to 3.5 kW (2013).

A possible time line is to upgrade LINAC by JFY2014 (400kW operation), and to upgrade MR by JFY2017 (750kW operation). Meanwhile, it is dependent on when the budget is secured to purchase MR magnet power supplies. It should be noted that, on May 2013, there was an accidental leakage of radioactive material outside of the hadron hall[9]. We are making our best effort to investigate the cause, and to prevent the recurrence by improving safety managements. Schedule for the beam operation and the upgrade plan has thus not yet been determined.

On the other hand, neutrino production target was designed for 750 kW operation with the beam energy of 30 GeV, $3.3 \times 10^{14}$ *ppp*, and 2.1sec cycle. The instantaneous temperature raise per each beam shot is to be 200 degree C and thermal shock is 7 MPa. It is to be compared to the tensile strength, 37 MPa, and safety factor is to be ~3.5, taking fatigue factor (0.9) into account. Oxidization to be caused by the contamination in helium gas reduces strength of graphite. Assuming 100 *ppm* of oxygen in helium gas, and temperature of graphite is kept to 700 degree C, the safety factor will be reduced to around 2 after 5 years of operation. We thus assume life of the target with original scenario is to be 5 years[6]. By adopting the doubled rep-rate scenario, number of protons per pulse was reduced to $2.2 \times 10^{14}$ *ppp*. Accordingly, instantaneous temperature rise / thermal shock per pulse will be reduced significantly, and the target may possibly accept beam with power even up to 1.5 MW.

Currently, the horns are being operated with 250 kA pulsed current and a rep-rate of 0.4 Hz. Two power supplies are used, one is for the horn-1 and the other is for the horn-2 and 3. To make doubled rep-rate scenario work, we will introduce individual power supply for each horn. This configuration requires less charging voltage, and thus reduces risk of failure for the power supplies. It will also make possible to operate horns with their rated current of 320 kA.

We have a plan to replace the entire target-horn system during 2013 summer shutdown. Based on the experiences to accept the 220 kW beam, various upgrades are being made for the new horns. These upgrades include a new port in each outer conductor for improved gas flow to the hydrogen-oxygen recombination system (this avoids possible explosion of hydrogen, which is produced by beam interacting with the cooling water), new low impedance strip lines, and improved cooling ducts as well as others[5]. The possible upgrade scenario for the other buildings and apparatuses in the neutrino experimental facility was presented in [10].